\documentstyle[12pt,epsfig]{article}
\begin{document}
{\hskip 12.0cm} SNUTP 96-095\par
{\hskip 12.0cm} hep-ph/9609410\par
\vspace{1ex}
{\hskip 12.0cm} January, 1997\\
\vspace{8ex}
\begin{center}        
{\LARGE \bf Phenomenology of A Supersymmetric Model for Fermion Mass 
Hierarchy}\\
\vspace{7ex}
{\sc Chun Liu}\footnote{\it email: liuc@ctp.snu.ac.kr}\\
\vspace{3ex}     
{\it Center For Theoretical Physics, Seoul National University}\\
{\it Seoul, 151-742, Korea}\\

\vspace{10.0ex}
{\large \bf Abstract}\\
\vspace{4ex}
\begin{minipage}{130mm}
                                                                               
   Some phenomenological aspects of a supersymmetric model for fermion 
mass hierarchy proposed previously are discussed.  It is required that
the lepton universality violation 
is near to its current experimental bound.  
The lepton number violation decay modes $\tau\rightarrow 2e \mu$ and 
$3\mu$ maybe observable in the near future.
The Majorana mass of electron-neutrino is predicted to be about 
$0.1$ eV.  The 
fine-tuning problem is discussed.\par
\vspace{0.5cm}
\end{minipage}
\end{center}

\newpage
                                                                               
   In this paper, we discuss some phenomenological aspects of a model
[1-3] which is proposed for understanding the fermion mass hierarchy
within the framework of low energy supersymmetry.  In the framework of 
R-parity violating supersymmetry [4], we thought that, in addition to 
the Yukawa interactions, the trilinear lepton number violating 
interactions also contribute masses to the fermions if the sneutrinos 
have nonvanishing vacuum expectation values (vevs).  We further introduced 
family symmetry among the three generations.  The sneutrino vevs make 
the family symmetry broken.  This may give a realistic pattern of fermion 
mass hierarchy.  CP violation occurs superweakly through sneutrino 
exchanges.  Although the model is interesting, it is necessary to discuss 
its detailed phenomenological implications.  We will show that the 
phenomenological discussions are indeed restrictive to the model.  Some
predictions will also be made.
\par
\vspace{1.0cm}    
   Let us first make a review of the model.  As for the particle contents,
besides all the fields of the minimal supersymmetric extension of the 
Standard Model, a SU(2)$\times$U(1) singlet superfield $X$ is introduced.
In addition to the gauge
symmetries which are the same as that in the Standard Model, one discrete
family  
symmetry is introduced.  It is a ${\rm Z}_3$ 
symmetry among the SU(2) doublets of the three generations.  Denoting $L_i$
and $Q_i$ as the SU(2) doublet superfields of leptons and quarks respectively, 
with $i=1, 2, 3$ standing for the three families, this ${\rm Z}_3$ symmetry 
says that the model is invariant under the cyclic operation 
$(L_1, Q_1)\rightarrow(L_2, Q_2)\rightarrow(L_3, Q_3)\rightarrow(L_1, Q_1)$.
Instead of the R-parity, the baryon number conservation is adopted. 
The superpotential then is  
\begin{equation}
\begin{array}{lll}
{\cal W}&=&\displaystyle g^l_j(\sum_{i}^{3}L_i^a)H_d^bE_j^c\epsilon_{ab}
+g^u_j(\sum_{i}^{3}Q_i^a)H_u^bU_j^c\epsilon_{ab}
+g^d_j(\sum_{i}^{3}Q_i^a)H_d^bD_j^c\epsilon_{ab}\\
&&+\lambda_j(L_1^aL_2^b+L_2^aL_3^b+L_3^aL_1^b)E_j^c\epsilon_{ab}\\
&&+\lambda^1_j(Q_1^aL_1^b+Q_2^aL_2^b+Q_3^aL_3^b)D^c_j\epsilon_{ab}\\
&&+\lambda^2_j(Q_2^aL_1^b+Q_3^aL_2^b+Q_1^aL_3^b)D^c_j\epsilon_{ab}\\
&&+\lambda^3_j(Q_3^aL_1^b+Q_1^aL_2^b+Q_2^aL_3^b)D^c_j\epsilon_{ab}\\
&&\displaystyle +\lambda' X(H^a_uH^b_d\epsilon_{ab}-\mu^2)~,
\end{array}
\end{equation}
with $a$ and $b$ being the SU(2) indices. 
And $E^c$, $U^c$ and $D^c$ stand for the SU(2) singlet antiparticle fields
of the leptons, up-type quarks and down-type quarks respectively; $H_u$ and
$H_d$ are the two Higgs doublets which couple to the $U^c$ field and $D^c$
field in Yukawa interactions respectively.
The term proportional to 
$\lambda'$ is nonvanishing so as to avoid a fourth massless neutrino [5]
and to break the gauge symmetries.
Actually these are the reasons for us to introduce the field $X$ [6]. 
\par
\vspace{1.0cm}
   The soft breaking terms should be added in the Lagrangian.  The 
principle in writing them is gauge invariance 
and baryon number conservation.  We did not assume exact
family symmetry in the soft breaking sector.  There may
be explicitly lepton number violating mass terms 
$\tilde{\mu}^2 \sum_{i} A_i^{\dag} \phi_d$ 
with $A_i$ and $\phi_d$ being the scalar 
components of $L_i$ and $H_d$ respectively.  So that there are no massless 
majoron as well as light scalar particles [6].
\par
\vspace{1.0cm}
   The scalar potential derived in Ref. [1] determines the vevs of the 
neutral scalar fields.  Besides Higgs fields, the scalar neutrinos also 
get nonzero vevs which are about $10-30$ GeV from the previous discussions 
[2, 3].
\par
\vspace{1.0cm}
   Given the model described above, a hierarchical pattern of fermion masses
can be obtained [1-3].  The charged lepton and 
quarks of the third generation 
get masses from the Yukawa interactions which actually give a kind
of democratic mixing of fermions.  The muon mass and the down quark mass 
originate from the tree-level trilinear R-parity violating interactions.
Whereas due to the soft breaking of the family symmetry, the electron, the 
charm quark and the strange quark 
might obtain masses from loops with neutralino 
and gluino internal lines respectively.  While the pattern of the mass 
hierarchy arises in an interesting manner in this model, some more careful 
discussions on the related phenomenology are needed.  Some general features 
of the R-parity violating models have been discussed already [7].  We are 
going to focus on some specific points of this model.
\par
\vspace{1.0cm}
{\it \bf 1. Lepton universality violation}\par
\vspace{0.5cm}
   Without loss of generality, we consider the scenario that only the third 
sneutrino field has nonvanishing vev which is denoted as $v_3$.  Such a 
scenario has been discussed elegantly by Ross and Valle [8].  Because 
$v_3\neq 0$, 
the fermions in superfield $L_3$ mix with that in Higgs and gauge superfields.  
For simplicity,
first we consider the case that the Yukawa and R-parity violating 
interactions are turned off.  For the fermion fields, we found in Ref. [1]
that the following composition which are orthogonal to the physical Higgsinos
and gauginos are still massless, 
\begin{equation}
\begin{array}{lll}
\nu'_3&=&\displaystyle N_{\nu'_3}(\nu_3-
\frac{v_3v_d}{v_d^2+v_u^2}\tilde{\phi}^0_d+\frac{v_3v_u}{v_d^2+v_u^2}
\tilde{\phi}^0_u)~,\\[3mm]
e'_3&=&\displaystyle N_{e'_3}(e_3-\frac{v_3}{v_d}\tilde{\phi}^-_d)~,
\end{array}
\end{equation}
with $N_{\nu'_3}$ and $N_{e'_3}$ being the normalization constants.
In Eq.(2), $(\nu_3,~ e_3)$ is the fermionic component of the superfield $L_3$.
$\tilde{\phi}_d$ and $\tilde{\phi}_u$ are the fermionic components of
the superfields $H_d$ and $H_u$ which have vevs $v_d$ and $v_u$ 
respectively.  
Besides the fermions $(\nu_1, e_1)$ and $(\nu_2, e_2)$ in the superfields
$L_1$ and $L_2$, $(\nu'_3, e'_3)$ is of course the third lepton doublet.
At this stage, all of the three lepton doublets are weak eigenstates.  Now
turn on the Yukawa interactions.  It can be shown [2] that the charged 
leptons
are still of democratic mixing [9]
which results in the mass eigenstate of the left-handed $\tau$,
\begin{equation}
\tau_L=\frac{1}{\sqrt{3}}(e_1+e_2+e'_3)~.
\end{equation}
At this stage, muon and electron are massless, so that the physical
eigenstates of them cannot be uniquely fixed.  They can be parameterized
as follows,
\begin{equation}
\begin{array}{lll}
\mu_L&=&\displaystyle\frac{1}{\sqrt{2}}(e_1-e_2)\cos\theta
+\frac{1}{\sqrt{6}}(e_1+e_2-2e'_3)\sin\theta~,\\[3mm]
e_L&=&\displaystyle -\frac{1}{\sqrt{2}}(e_1-e_2)\sin\theta
+\frac{1}{\sqrt{6}}(e_1+e_2-2e'_3)\cos\theta~,\\[3mm]
\end{array}
\end{equation}
where $\theta$ cannot be determined until the muon mass basis is fixed.
\par
\vspace{1.0cm}
   After the involvement of R-parity violation, muon gets mass.
From Eq. (1) we see that in terms of the mass eigenstates ($e_L$, 
$\mu_L$, 
$\tau_L$), the relevant part for the leptons in the superpotential is 
\begin{equation}
{\cal W}'=g^{\tau}L_{\tau}^aH_d^bE_{\tau}^c\epsilon_{ab}+
L_eL_{\mu}(\lambda_{\mu}E_{\mu}^c+\lambda_{\tau}E_{\tau}^c)~.
\end{equation}
$E_{\tau}^c$ is propotional to $g_j^lE_j^c$. 
$g^{\tau}$ and $\lambda_{\mu}$, $\lambda_{\tau}$ are the Yukawa
coupling constant and the R-parity violating coupling constants in the mass
eigenstates, respectively.  They are compositions of $g^l_j$, $\lambda_j$ 
($j=1, ~2, ~3$) in Eq. (1).  $E^c_{\mu}$ is a combination of $E^c_j$,
which is orthogonal to $E^c_{\tau}$.  It will be proved to be the
right-handed muon mass eigenstate.  Up to now, it is not neccessarily the
mass eigenstate, however we always have the freedom to write the 
superpotential in the above form.  Assumption that only $\nu_3$ gets a vev
leads that $\cos\theta=1$, and it is at this stage we see $E^c_{\mu}$
is just (approximately) corresponding to the mass eigenstate of the 
right-handed muon. 
\par
\vspace{1.0cm}
   It is because $\nu'_3$ and $e'_3$ in Eq. (2) do not coincide
in form that makes the leptons deviate from their universality.  Compared to 
the $e_1\rightarrow \nu_1$ weak transition, the $e'_3\rightarrow \nu'_3$
transition amplitude is suppressed by a factor of 
$C=N_{\nu'_3}N_{e'_3}(1+\frac{v_3^2}{v_d^2+v_u^2})$.  
For $e\rightarrow \nu_e$ weak transition, where 
$\nu_e\equiv \frac{1}{\sqrt{6}}(\nu_1+\nu_2-2\nu'_3)$, the suppression
factor is $\tilde{C}_e=\frac{1}{3}(1+2C)$ which can be 
effectively absorbed into the gauge interaction coupling constant $g^e$.
For $\mu\rightarrow\nu_{\mu}$ weak transition, due to ($\nu'_3, ~e'_3$)
is not the composition, the corresponding factor $\tilde{C}_{\mu}=1$.
For the 
$\tau\rightarrow \nu_{\tau}$ weak decay, where 
$\nu_{\tau}=\frac{1}{\sqrt{3}}(\nu_1+\nu_2+\nu'_3)$, the suppression factor
is $\tilde{C}_{\tau}=\frac{1}{3}(2+C)$ which can be effectively absorbed 
into the
gauge interaction 
coupling constant $g^{\tau}$.  Therefore 
$\tilde{C}_l\neq 1~(l=e, ~\tau)$ just 
measures the lepton universality violation.
With reasonable choice of $\tan \beta$, like $\tan \beta \simeq 2.2$ [3], 
if $v_3$ were taken as $30$ GeV, the violation of factor 
$C$ from unity would be
as large as $10\%$.  Only if $v_3\leq 10$ GeV, can this violation be
within $0.4\%$,
and hence the violation of the factors $\tilde{C}_l$ from unity smaller 
enough to satisfy the current experiment [10].  However,
the value of $v_3$ cannot be too small in this model, because it contributes
muon mass.  We will take $v_3$ as  $10$ GeV in the following.
With such choice of parameters, we have the following lepton universality
violation, 
\begin{equation}
\begin{array}{lll}
g^e:g^{\mu}:g^{\tau}&=&\tilde{C}_e:\tilde{C}_{\mu}:\tilde{C}_{\tau}
=\frac{1}{3}(1+2C):1:\frac{1}{3}(2+C)\\
&=&0.997:1:0.999=1:1.003:1.002~,
\end{array}
\end{equation}
which is still consistent with experiments [10].
\par
\vspace{1.0cm}
   It should be remarked here that the above discussion on the lepton 
universality violation is different from that in Refs. [5] and [8].  We have 
introduced the coupling $\lambda'$ in this model (see Eq. (1)), there is 
no fourth massless neutrino.  So the violation cannot be rotated away by 
redefining the $\tau$-neutrino field.
\par
\vspace{1.0cm}
{\it \bf 2. Lepton number violation and FCNC}\par
\vspace{0.5cm}
   Generally this model allows for 
lepton number violation processes at tree level.  The violation is  
caused by the trilinear R-parity violating terms with the slepton
exchange.  From Eq. (5), it is interesting to see that the decay 
$\mu\rightarrow 3e$ does not occur in this model.  Therefore, this model
avoids one of the most stringent restriction of the lepton number
violation.  Lepton number violation occurs in $\tau$ rare decays.
If we consider the processes with only charged leptons as final states,
the decay modes $\tau\rightarrow 2e\mu$ and $\tau\rightarrow 3\mu$ can 
occur, as shown explicitly in Figure 1.  They have the 
same decay rate,
\begin{equation}
\Gamma=\displaystyle 
\frac{[\frac{\lambda_{\mu}\lambda_{\tau}}
{m_{\tilde{\nu}}^2}]^2m_{\tau}^5}{192\pi^3}~.
\end{equation}
From the muon mass 
$m_{\mu}\simeq \lambda_{\mu}v_3$, we get 
$\lambda_{\mu}\sim 10^{-2}$.  It is reasonable to assume that the value
of $\lambda_{\tau}$ has the same order of magnitude as $\lambda_{\mu}$.
In this case, the branching ratio of  
the decays 
$\tau\rightarrow 2e \mu$ and $3\mu$ are
$10^{-7} - 10^{-11}$ if the sneutrino mass is 
around $100 - 1000$ GeV.  Such a branching ratio maybe observable in the 
near future.  The modes $\tau\rightarrow 3e$ or $2\mu e$, however cannot
occur in this model.
\par
\vspace{1.0cm}
   A remark should be made on the slepton SUSY breaking parameters.  The 
lepton flavor violation, like in the proccess $\mu\rightarrow e \gamma$, 
can be induced by the SUSY breaking parameters through loops.  The 
experimental data on such process put severe constraints on these 
parameters.  The symmetries assumed in this model, however, do not 
constrain these parameters strong enough.  Similar to what have been done 
for the quark case [3], further assumptions are needed here.  They are
(i) the slepton mass-squared matrice corresponding to definite chiralities
of leptons are proportional to unit matrix; (ii) the slepton mass-squared 
matrix corresponding to the mixing associated with different chiralities
is propotional to the lepton mass matrix.  
\par
\vspace{1.0cm}
   However, these conditions cast doubt on the electron and neutrino mass
generation mechanism through loops proposed by Ref. [2].  In that mechanism,
we hoped to violate
condition (i) to give electron mass through loop.
But the violation is required
to be small and can be measured by some dimensionless quantities $\delta$,
that are defined as square of the ratios of the flavor off-diagonal masses
to the average slepton mass.  From the experimental constraints [11],
$\delta\leq (10^{-2}-10^{-3})\times (\frac{m_{\tilde{l}({\rm GeV})}}{100})^2$, 
when the averaged slepton mass
$m_{\tilde{l}}$ is compatible with the photino mass $m_{\tilde{\gamma}}$.
According to Ref. [2], the electron mass should be 
$m_e\simeq \frac{\alpha}{4\pi}\delta\frac{\tilde{m}v_d}{m_{\tilde{l}}}$,
where $\tilde{m}$ is soft mass parameter associated with different
chirality mixing, which might give realistic numerical 
result of electron mass, if
$\tilde{m}\simeq 1$ TeV were taken.  Such a value of $\tilde{m}$ is too
large to be acceptable.
\par
\vspace{1.0cm}
{\it \bf 3. Neutrino masses}\par
\vspace{0.5cm}
   The loop induced Majorana neutrino mass is inevitable in this model.
The best way in studying this problem is to work in the basis of the
mass eigenstate.  
The mechanism for the loop-level neutrino
mass was discussed in Ref. [12].  
From Eq. (5), we see that only electron type neutrino can be massive 
through loop as shown in Figure 2, the induced mass is
\begin{equation}
\begin{array}{lll}
m_{\nu_e}&\simeq&\displaystyle\frac{(\lambda_{\mu})^2}{16\pi^2}
\frac{\tilde{m}^0m_{\mu}^2}{m_{\tilde{l}}^2}~,\\[3mm]
&\simeq&\displaystyle\frac{(\lambda_{\mu})^2}{16\pi^2}
\frac{\tilde{m}v_d}{m_{\tilde{l}}^2}\frac{m_{\mu}^2}{m_{\tau}}~.
\end{array}
\end{equation} 
In above equation, we reexpressed the parameter $\tilde{m}$ by
$\tilde{m}^0$ which is a more natural choice in the lepton case.  For
reasonable value of $\tilde{m}^0$ and $m_{\tilde{l}}$, 
$\tilde{m}^0\simeq m_{\tilde{l}}\simeq 100$
GeV, we predict Majorana $\nu_e$ mass is around 0.1 eV.  
Note that only muon and its superpartner contribute the $\nu_e$ mass
in the loop.  At this stage, the model predicts vanishing masses of
$\nu_{\mu}$ and $\nu_{\tau}$.  And there is no mixing 
between $\nu_e$ and them.
\par
\vspace{1.0cm}
   It is necessary to discuss the tree-level induced neutrino
mass.  Usually in the case of large sneutrino vev, we expect that,
because of the mixing with Zino, neutrino gets a large mass by see-saw
mechanism unless an unnatural fine-tuning is made.  But this model
avoids such tree-level neutrino mass, as shown explicitly in the 
following.  The Lagrangian for the neutralino masses is given as
\begin{equation}
-i(\nu_3~~ \tilde{\phi}^0_d~~ \tilde{\phi}^0_u~~ \tilde{Z}~~ \tilde{X})
\left(\begin{array}{ccccc}
0   &0          &0            &av_3         &0           \\
0   &0          &0            &av_d         &\lambda' v_u\\
0   &0          &0            &-av_u        &\lambda' v_d\\
av_3&av_d       &-av_u        &M_{\tilde{Z}}&0           \\
0   &\lambda'v_u&\lambda'v_d  &0            &0       
\end{array}
\right)
\left(\begin{array}{c}
\nu_3\\ \tilde{\phi}^0_d\\ \tilde{\phi}^0_u \\ \tilde{Z}\\ \tilde{X}
\end{array}
\right)
+{\rm h.c.},
\end{equation}
where $a=(\frac{g^2+g'^2}{2})^{1/2}$ with $g$ and $g'$ the 
SU(2)$\times$U(1) gauge coupling constants.  Note that 
the Zino mass $M_{\tilde{Z}}$ of the soft breaking term has also been
included in.  It is easy to see that
the above mass matrix is of rank 4 (instead of 5) with eigenstate
$\nu'_3$ expressed in Eq. (2) corresponding to the zero 
eigenvalue.  However, this does not mean that there is no fine-tuning
problem in this model.  This problem has actually transferred to the
smallness of the coupling constants of the terms like
$H_uH_d$, $H_uL_i$ and $XH_uL_i$ which we have not written in Eq. (1).
While the smallness needs further
explanation, less fine-tuning is needed in this model
because of supersymmetry.
\par
\vspace{1.0cm}
   In summary, the supersymmetric model for fermion mass hierarchy 
proposed in Refs. [1-3] has got restrictive limitations from 
phenomenological considerations.  It requires that the 
lepton universality violation is near to
its experimental bound.  
The lepton number violation decay modes $\tau\rightarrow 2e \mu$ and 
$3\mu$ maybe observable in the future's experiments.
The electron neutrino Majorana mass is predicted to be about 0.1 eV.
\par
\vspace{1.0cm}
   Finally a remark should be made on the electron mass.  This model
gives an interesting hierarchical pattern for leptons.  However the idea
of radiative generation for electron mass [2] is seriously problematic.  
Actually such kind of idea does not make simplification torwards the
understanding of the lepton mass hierarchy, instead it merely transfers
the problem from the Yukawa sector to other sector of the model.  
Therefore it is fair to say that the idea of radiative mass generation
is not appealing in the framework of this model.  It is preferable if
the electron mass originates in some new physics which has not yet been 
included in the Lagrangian.
\par
\vspace{2.0cm}

   The author would like to express his sincere gratitude to the referee
for his (her) many constructive suggestions, and also 
thank many colleagues in Beijing, as well as 
H. Sato for helpful discussions. This work was supported in part by 
KOSEF through SRC.

\newpage
\noindent
\begin{center} {\Large\bf FIGURE CAPTIONS} \end{center}
\vskip1cm
Fig. 1
\hskip .3cm
Feynman diagrams for the decays $\tau\rightarrow 2e \mu$ and 
$3\mu$.
\\
Fig. 2
\hskip .3cm
The mechanism for electron-neutrino mass generation.
\newpage
\pagestyle{empty}
\begin{center} {\Large\bf FIGURES} \end{center}
\vskip .5cm
\hbox to\textwidth{\hss\epsfig{file=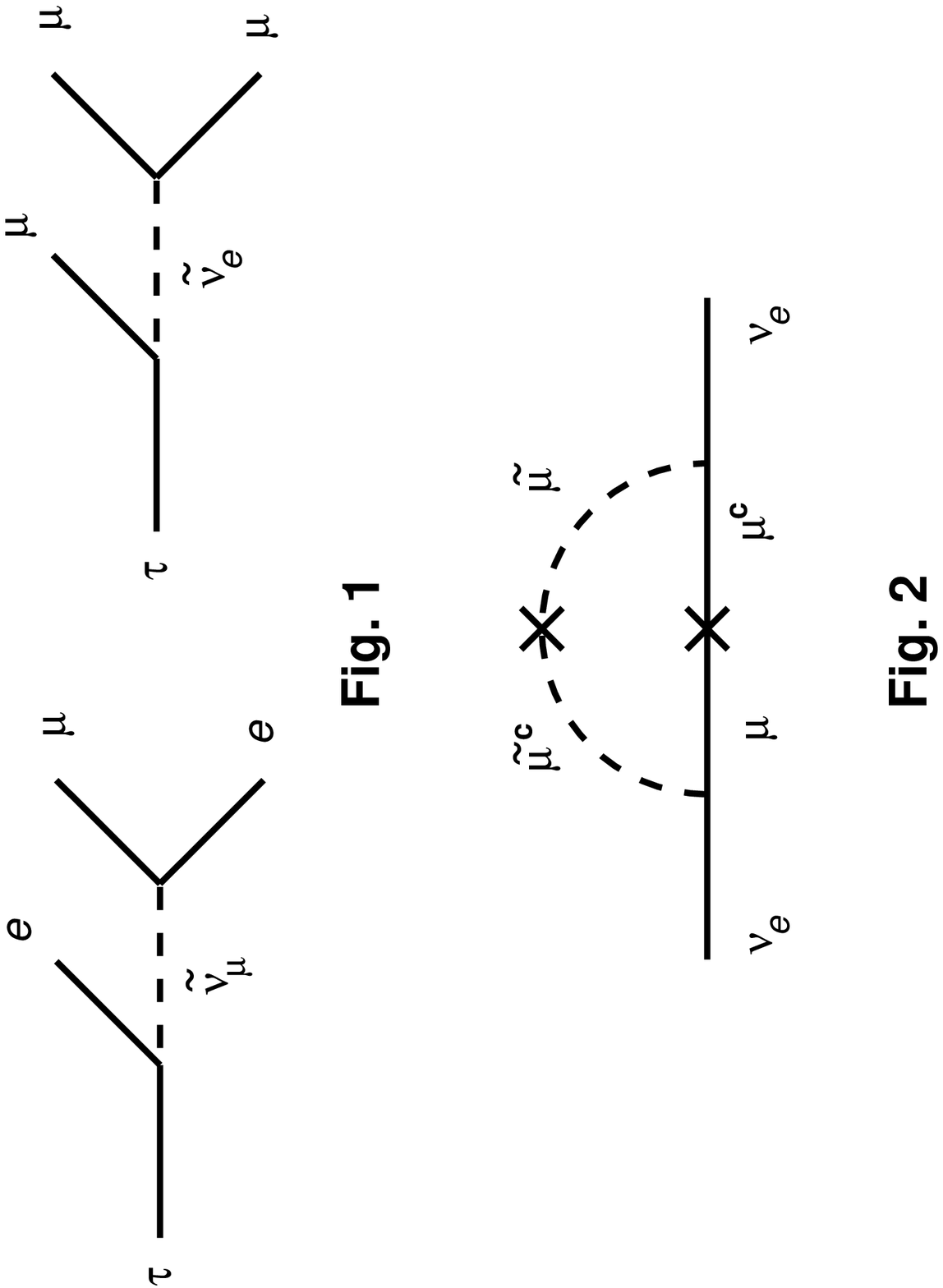,height=17cm}\hss}
\end{document}